\documentclass[letter]{aa} 
\usepackage{graphicx}
\usepackage{txfonts}
\usepackage{epsfig}
\usepackage{amsmath} 
\usepackage{rotating}           
\usepackage{color}     
\usepackage{graphicx}
\usepackage{upgreek} 

\def\mcube {$\hbox{{m}}^{-3}$} 
\def\kms{km~s$^{-1}$}
\def\ms{m~s$^{-1}$}
\def\mss{m~s$^{-2}$}

\def\degC{$^{\circ}$C}

\def\lapp{\ifmmode\stackrel{<}{_{\sim}}\else$\stackrel{<}{_{\sim}}$\fi}
\def\gapp{\ifmmode\stackrel{>}{_{\sim}}\else$\stackrel{>}{_{\sim}}$\fi}

\begin{document} 
\title{`Oumuamua as a light sail -- evidence against artificial origin}
 \author{S. J. Curran}

  \institute{School of Chemical and Physical Sciences, Victoria University of Wellington, PO Box 600, Wellington 6140, New Zealand\\
  \email{Stephen.Curran@vuw.ac.nz} 
 }
 
\abstract{ 
`Oumuamua, the first detected interstellar visitor to the solar system, exhibits non-gravitational acceleration in its
  trajectory. Ruling out other means of propulsion, such as the evaporation of material via a cometary tail, 
it has been argued that radiation pressure is responsible for this acceleration. From this ($a=5\times10^{-6}$ \mss), the mass
  of the object must be approximately $4\times10^4$~kg, and given its dimensions, `Oumuamua must have a thickness of
  $\lapp1$~mm if of a similar rock/iron composition as the Earth.  
This raises the much publicised possibility that `Oumuamua is
  artificial in origin, sent intentionally across interstellar space by an alien civilisation, 
This conclusion,
  however, relies upon the common misapprehension that light (solar) sails can accelerate to a considerable fraction of
  the speed of light, permitting rapid interstellar travel. We show that such speeds are unattainable for conceptual
  man-made sails and that, based upon its observed parameters, `Oumuamua would require half a billion years just to
  travel to our solar system from its closest likely system of origin.  These cosmological time-scales
make it very unlikely that this is a probe sent by an alien civilisation.
}
  

   \maketitle
%

\section{Introduction} 
\label{intro}

`Oumuamua (1I/2017 U1) was discovered on 19 October 2017 \citep{mwm+17} by the {\em Panoramic Survey Telescope and Rapid
  Response System 1} (Pan-STARRS1) survey \citep{djg+13,wcl+16}. The highly hyperbolic trajectory, with a speed of 26
\kms, indicated that the object originated from outside of the solar system in the direction of Lyra \citep{mwm+17}. As
the first detection of a visitor unbound by the Sun's gravity, `Oumuamua has generated much interest, and much
controversy -- specifically over the origin of its $5\times10^{-6}$~\mss\ non-gravitational acceleration
\citep{mfm+18}. While this has been disputed (e.g.  \citealt{kat19}), 
ruling out the usual suspects, such as thrust from a cometary tail, it has been suggested that
radiation pressure drives the acceleration \citep{bl18}.  That is, `Oumuamua is a light sail.

Over the past century, radiation pressure has been proposed as
a means to propel a payload through space \citep{zan25,for84,fse16}. Indeed, the principle was put into practice when
using the solar panels to correct the trajectory of Mariner 10 during its flyby of Mercury in 1974.
The advantage of a light sail is that it is powered by an external source, such as a star, releasing it from the burden
of an onboard fuel supply, which is the main disadvantage of a rocket.  This necessity is further compounded by the rocket
needing fuel not just to accelerate the payload, but the mass of the fuel itself.  Thus, light sails possibly offer a
more practical means of exploring deep space, with accelerations to relativistic velocities often suggested,
allowing us to reach the nearest extrasolar star, Proxima Centauri (at 4.22 light years distance), in a matter of
decades (e.g. \citealt{lub16,pop17,wdk18,ll20}).  

Here we show that, even if we neglect slowing (and damage) by interplanetary material, there exists an
effective  terminal velocity beyond which the sail barely accelerates. This velocity is much lower than
the relativistic speeds proposed for  conceptual light sails (e.g. \citealt{kip17}), meaning that the 
travel times are vastly underestimated. Using the 
properties of `Oumuamua, we find  the terminal velocity to be $\lapp1$~\kms, thus requiring, at the very least, millions
of years for interstellar travel.

\section{Analysis}

\subsection{Acceleration by light}

The intensity of radiation on an object at distance $r$ from a source of luminosity $L$ is
$I = {L}/({4\pi r^2})$,
with the power intercepted over a projected facing surface area, $A_{\text{eff}}$, being $P=IA_{\text{eff}}$. 
The energy carried by each photon is $E = hc/\lambda$ and so the number of photons intercepted each second is
\[
n = \frac{IA_{\text{eff}}}{E} = \frac{IA_{\text{eff}} \lambda}{hc}.
\]
The momentum carried by each photon is $p = h/\lambda$, of which $2bp\cos\theta$ is imparted to the sail, where the 
factor of two conserves the momentum for a perfectly reflected photon. $b$ is the albedo 
and $\theta$ the angle the photon strikes with respect to the normal to the sail.
For $\theta = 0$, over the whole sail, the total momentum imparted each second is therefore
\[
\dot{p}_{\text{t}}\equiv \dot{p} n = 2b\frac{h}{\lambda}\frac{IA_{\text{eff}}\lambda}{hc} = \frac{2bIA_{\text{eff}}}{c}.
\]
Since the force $F \equiv {dp_{\text{t}}}/{dt}$, in the non-relativistic regime the acceleration 
is given by
\begin{equation}
a= \frac{dp_{\text{t}}}{mdt}  = \frac{2bIA_{\text{eff}}}{mc} = \frac{bLA_{\text{eff}}}{2\pi mc r^2},
\label{eq1}
\end{equation}
where $m$ is the total mass (sail plus payload).

\subsection{Interstellar travel by conceptual light sails}

\subsubsection{Terminal velocity}

From the sail properties (Equ.~\ref{eq1}), 
the acceleration is maximised by maximising the albedo and area and minimising the mass.
For example, a theoretical sail of size 1~km on a side,
constructed from lithium (the lightest metal with a density of $\rho = 530$~kg~m$^{-3}$, \citealt{wri92}) and of mass 1~kg
would have a thickness of 1.8~nm. This constitutes a sail thickness  only a few atoms wide, which must remain flat
and stiff over its million square metre area, while surviving the rigours of space travel. 
Neglecting the considerable challenge in materials science and engineering required, if launched from the Earth's
orbit, this would have an initial acceleration of 8.17~\mss\ ($0.83g$) and, if sustained,
 would give a travel time of just 2.2~years to Proxima Centauri. 

\begin{figure}
\centering \includegraphics[angle=-90,scale=0.50]{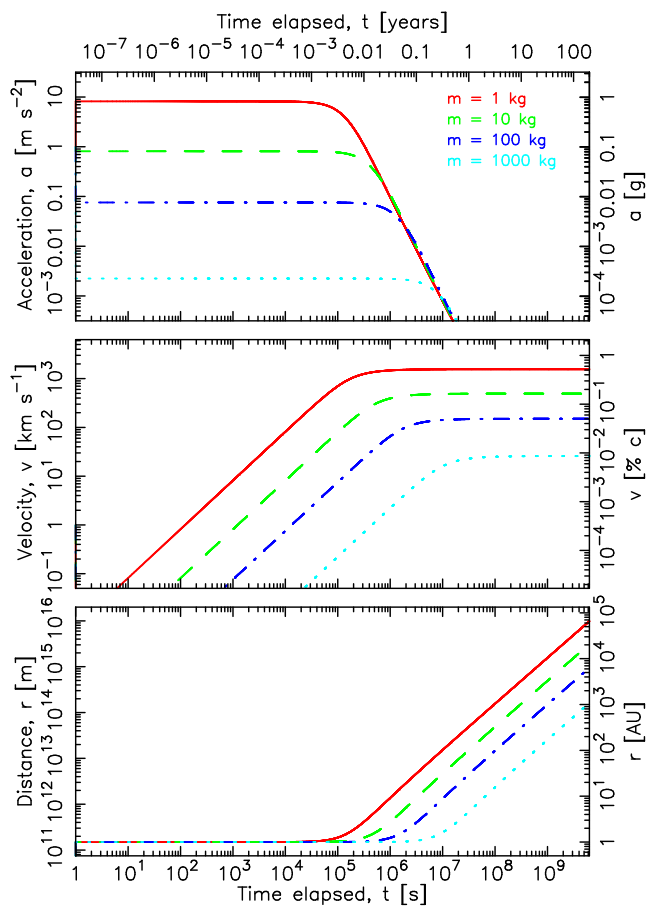}
\caption{Acceleration (top), velocity (middle) and distance (bottom panel) travelled by  
a conceptual light sail  ($A_{\text{eff}}= 10^6$~m$^2$, $b = 0.9$) for different masses starting at 1 AU (149\,597\,871 km) from the Sun.}
\label{f1}
\end{figure}

However, many conceptual light powered journeys do not consider that a large acceleration
removes you from the  illuminating power source quicker, causing a large decrease in further acceleration,
This is described by the second order ordinary differential equation
\begin{equation}
\frac{d^2r}{dt^2} = \frac{bLA_{\text{eff}}}{2\pi mc r^2}.
\label{eq2}
\end{equation}
Solving this numerically, using the equations of motion over small time ranges (from $\Delta t = 10^{-9}$~s, 
depending upon the regime), we find that, while the 
acceleration asymptotically approaches  zero never quite reaching it, it rapidly gets so small 
as to take cosmological time-scales to increase the speed
by an additional metre per second. 
This introduces an effective {\em terminal velocity}, which in the case of the
conceptual sail above is 1563~\kms\ ($0.0052c$)\footnote{Modelled from $a=8.17$ to  $9.9\times10^{-19}$~\mss.}, thus taking the sail 810 years to reach Proxima Centauri (Fig.~\ref{f1}).\footnote{Either by passing at this speed or using Proxima Centauri to decelerate around the half way mark, assuming a solar luminosity.}
While a terminal velocity has been discussed previously \citep{kip17}, this occurs  in the relativistic regime. 

\subsubsection{Maximising the acceleration}
\label{mta}

From Equ.~\ref{eq1}, the acceleration can be increased by increasing the sail area and albedo or by decreasing the
distance to the power source or the total mass.  Increasing the area and albedo are clearly challenging, although 
the distance can be decreased by a swing-by around the
Sun \citep{clg15}. 

\begin{figure}
\centering \includegraphics[angle=-90,scale=0.50]{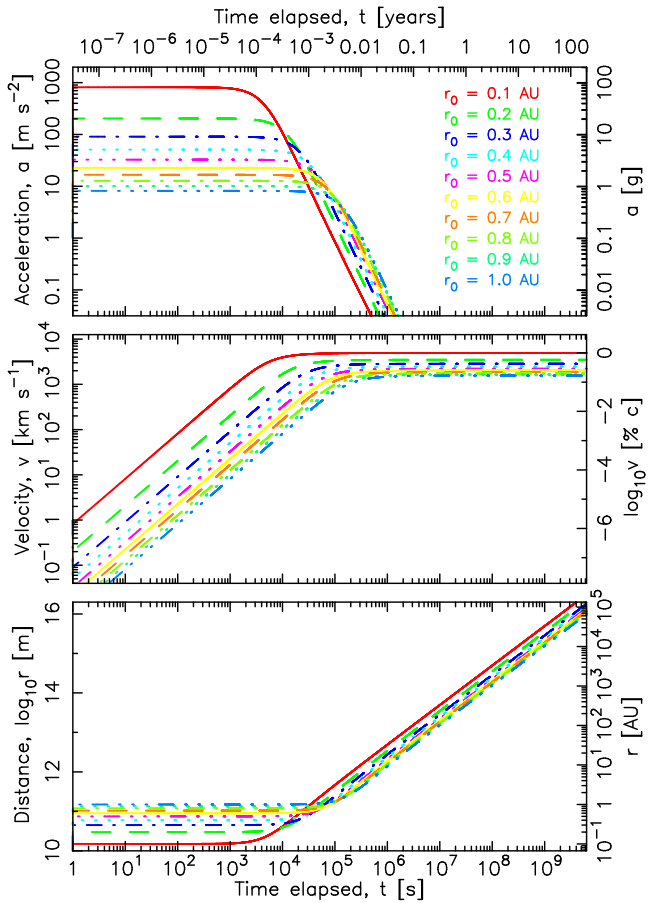}
\caption{Acceleration (top), velocity (middle) and distance (bottom panel) travelled by  
a conceptual light sail  ($A_{\text{eff}}= 10^6$~m$^2$, $b = 0.9$) of total mass 1~kg at different launch distances from the Sun.}
\label{f2}
\end{figure}
\begin{figure}
\centering \includegraphics[angle=-90,scale=0.5]{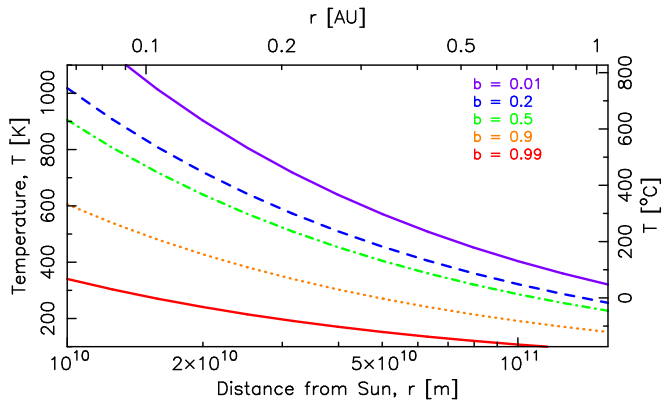}
\caption{Temperature of a black body as a function of  distance from the Sun for different albedos.}
\label{f3}
\end{figure}

 
For example, a million square metre sail of total mass 1~kg launched 0.1~AU from the Sun\footnote{Within Mercury's orbit
  of 0.44 AU.}, would have an initial acceleration of 817~\mss\ and a terminal velocity of 4943 \kms\ ($0.016c$),
reducing the journey to Proxima Centauri to 260 years (Fig.~\ref{f2}).  In addition to the loads on this conceptual
sail, the thermal properties must be considered, given that, even for the reflectivities discussed for
concept sails ($b=0.9$), for a thin sail\footnote{$A_{\text{surface}} \approx 2A_{\text{eff}}$.} in thermal equilibrium temperatures would be in excess of 200\degC\ at 0.1 AU
(Fig.~\ref{f3}).

Other materials, such as polyethylene terephthalate (mylar) and carbon fibre, have been proposed in order to lighten 
conceptual sails \citep{lan99,lan03a}. Both are, however, denser than lithium ($\rho = 1400$ and $1800$~kg~m$^{-3}$, respectively),
although the latter, if porous,  can be as low as $\rho = 270$~kg~m$^{-3}$ \citep{sg03}. 
However, this is
only about half the density of lithium,  resulting in similar sail widths  while not being as strong, 
with a tensile strength of $1.7$~MPa,
compared to $15$~MPa for lithium. Furthermore, carbon fibre would
require a reflective coating, further increasing the mass, to have a similar albedo as  a metal.

We can conceptualise the fastest ``possible'' theoretical solar driven sail, by assuming the ability to construct and deploy
a  sail of arbitrarily large size with a thickness of one atom and perfect reflectivity ($b=1$). The only
constraint is the minimum mass dictated by the volume of the sail, for which we neglect the mass
of the payload and other ancillary requirements (such as strengthening, stiffening and unfurling mechanisms).
We choose the material to be lithium (diameter $3$~\AA) since this is only slightly denser than a carbon fibre foam,
which, by definition, cannot form a structure only one atom thick.
Since the area and mass are degenerate, $bA_{\text{eff}}/m < 6.086\times10^6$~m$^2$~kg$^{-1}$
no matter the area. 
Therefore, the
ideal sail gives, at best, $v_{\text{term}} = 4066$~\kms\ ($0.014c$), which results in 
a $>310$ year journey to Proxima Centauri. As discussed above, the acceleration can be increased by
launching closer to the Sun.
\begin{figure}
\centering \includegraphics[angle=-90,scale=0.5]{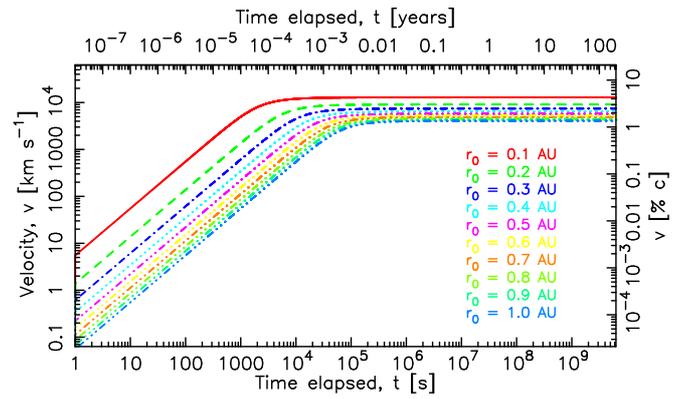}
\caption{Velocity of  the best possible theoretical sail at different launch distances from the Sun.}
\label{fb}
\end{figure} 
For a launch at 0.1 AU, the initial acceleration is $5.53\times10^3$~\mss\ ($564g$), and the
terminal velocity  $v_{\text{term}} = 1.29\times10^4$~\kms\ ($0.043c$, Fig.~\ref{fb}), thus
taking 98 years to reach  Proxima Centauri. 

The conceptual project {\em Breakthrough Starshot} proposes taking advantage of the high initial acceleration by using
100 GW of Earth-based laser power for several minutes only, accelerating a 
4 metre wide sail of 2.4~g mass to 
$0.2c$ over this time, reaching Proxima Centauri in 20 years \citep{lub16}. 
However, not only is such laser power a million times more powerful than current continuous
lasers, for the lowest densities discussed above
($\rho = 270$~kg~m$^{-3}$ for porous carbon fibre), the width of the sail remains incredibly thin at $0.5~\mu$m,
 {\em without} additional payload, which must survive an initial acceleration 
of $7\times10^5$~\mss\ ($80\,000g$). Other fundamental technical challenges include widening of the laser beam
(dictated by diffraction), reducing the incident power, and vaporisation of the sail by the laser \citep{kip17,kat21}.

\subsection{`Oumuamua as a light sail}

If the acceleration of `Oumuamua is driven by radiation pressure,
it qualifies as a light sail. Furthermore, if this is the sole source of its acceleration, from its measured area
the object must
be relatively thin, thus the motivation for proposing an
artificial origin \citep{bl18,loe18a,loe18,wil18,loe21}. Here we examine the feasibility of this suggestion by comparing 
the implications of it being a light sail with the conceptual examples discussed above. 

The dimensions of  `Oumuamua  are degenerate with its albedo 
(Table~\ref{t1}),
\begin{table}
\centering
  \caption{Effective diameter and albedo of `Oumuamua \citep{tmh+18,bbd+19}.}
\begin{tabular}{ll c c c} 
\hline\hline
\smallskip
Albedo, $b$ &  $d_{\text{eff}}$ [m] & $A_{\text{eff}}$ [m$^2$] &  $b A_{\text{eff}}$ [m$^2$] \\
\hline
$>0.2$ & $<98$ & $<7543$ &  1509\\
$>0.1$ & $<140$ & $<15\,400$ & 1540 \\
$>0.01$ & $<440$ & $<152\,000$ &  1520 \\
\hline
\end{tabular}
\label{t1}  
\end{table} 
Given that the 
acceleration  was $4.9\pm0.2\times10^{-6}$~\mss\ when it was at 
0.2556 AU from the Sun \citep{mfm+18},
results in a mass of $42\times10^3$~kg. If of a rock/iron
composition, similar to the Earth ($\overline{\rho} = 5500$ ~kg~\mcube), the thickness is
somewhere between 50~$\mu$m and 1~mm, depending upon the actual area (Table~\ref{t1}). The
former is of the order of the width of a human hair, although this range can be increased from 1~mm
to 2~cm if constructed of a more exotic material such as porous carbon fibre. 

\begin{figure}
\centering \includegraphics[angle=-90,scale=0.5]{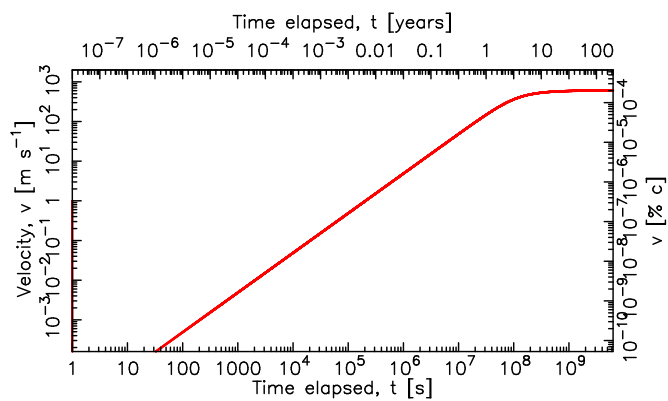}
\caption{Velocity of `Oumuamua based upon its observed acceleration.}
\label{fo}
\end{figure}

From the observed parameters of `Oumuamua \citep{tmh+18,bbd+19}, we determine a terminal velocity of  $610$~\ms\ (Fig.~\ref{fo}),  at which it would take $2\times10^6$~years to reach Proxima Centauri.  
The acceleration can be increased by launching from closer to the parent star and at 0.1 AU from a star of solar
luminosity, the acceleration increases by an order of magnitude to $3.2\times10^{-5}$~\mss, giving a terminal velocity of
close to 1 \kms, although, from the range of possible albedos ($b = 0.01-0.2$,
Table~\ref{t1}), the temperature is high at 540--750\degC . 
If sent by an alien civilisation, we can envisage exotic materials, where close to 100\% reflectivity
($b\approx1$) is possible, allowing launches from arbitrarily close to the star, giving a much larger kick in
acceleration. However, not only is such a large albedo not observed for `Oumuamua, but, as seen from Fig.~\ref{f2}, the
benefits from a large initial acceleration are tempered by a more rapid distancing from the power source, thus not
yielding a proportionate return for the effort.  In addition to the incident flux, the acceleration is limited by the
area of the sail and the mass (Equ.~\ref{eq1}).  For `Oumuamua the former is constrained by observation (Table~\ref{t1})
and the latter on the presumption that the acceleration is due to radiation pressure. If working under this presumption,
from the travel times (of the order a million years to the nearest
star), 
it is clear that this is much inferior to the light sails being conceived by ourselves.

\section{Discussion and conclusion}

We have shown that, based on its observed parameters, `Oumuamua would take two million years to cover the distance to
the nearest extrasolar star (at just 4.22 light years distance). This is a vast time scale, even in comparison to
conceptual man-made light sails, for which we show the estimated travel times to be vastly optimistic.  Based upon its
trajectory, a more distant ($\approx200$ light years) point of origin is proposed \citep{gwk17}, suggesting a travel
time of the order of half a billion years.

From its measured velocity of 26 \kms\ \citep{mwm+17}, it would take `Oumuamua 50\,000 years to travel from Proxima Centauri and
400\,000 years to travel 10~pc \citep{zuc21}, encompassing the nearest 357 main sequence stars \citep{hjw+18}. However,
we have shown that such a velocity cannot be achieved by `Oumuamua as a solar sail, with the measured speed being due to
the Sun's gravitational attraction.  Even at a  travel time of 400\,000 years, it has been argued that an advanced
civilisation would use its resources to explore interstellar space via electromagnetic waves, rather than the
``construction and launch of an `Oumuamua-like probe'' \citep{zuc21}.

There is of course the possibility of using lasers to accelerate a sail to relativistic speeds in the space of a few
minutes, as proposed for {\em Breakthrough Starshot}. However, even for a mass of just 2.4~g and a theoretical
albedo of $b>0.9999$, which no metal has, even in the microwave band, this is fraught with many, perhaps insurmountable,
challenges \citep{kat21}. `Oumuamua has a much lower, and realistic, albedo and even if we assume the possibility that
the surface has been tarnished from $b\approx1$ by its journey, its mass implies a power requirement of $\sim10^{18}$~W,
which is $\sim10^{13}$ times today's most powerful lasers.  
Other fantastic power sources include massive stars, microquasars, supernovae, pulsars and active
galactic nuclei \citep{ll20}.  However, life cannot evolve in proximity to any of these objects and there is no
discussion of how the sail would be transported close enough in order
to take advantage of their immense power output. Thus, the Sun (or any parent star) remains the best option,
since it provides vast amounts of continuous power for free, the full utilisation of which would be an indicator of an
advanced civilisation \citep{dys60}.

It has been suggested that `Oumuamua is in fact accelerated by out-gassing \citep{mfm+18}, 
but even if this is ruled out \citep{ll20, sl21},  so that `Oumuamua is indeed a light sail \citep{bl18}, given the likely
cosmological time scales required to traverse between stars, we conclude that it is unlikely that `Oumuamua has been sent by an
extraterrestrial civilisation and more likely that it is just an unusually shaped rock, which has happened to wander into the
solar system.

\section*{Acknowledgements}

I would like to thank the anonymous referee for their helpful comments.
This research has made use of NASA's Astrophysics Data System Bibliographic Service.

\end{document}